\documentclass[a4paper]{jpconf}
\usepackage{graphicx}
\newcommand{\solm}{M$_{\odot}$\ }

\begin{document}
\title{Flare emission from Sagittarius A*}

\author{A. Eckart$^{1,2}$,
      M. Garc\'{\i}a-Mar\'{\i}n$^1$,
      S. N. Vogel$^{3}$,
      P. Teuben$^{3}$,
      M.R. Morris$^4$, 
      F. Baganoff$^5$,
      J. Dexter$^{6,7}$,
      R. Sch\"odel$^{8}$,\\
      G. Witzel$^1$,
      M. Valencia-S.$^{1,2}$,
      V. Karas$^9$,   
      D. Kunneriath$^9$,   
      M. Bremer$^1$,
      C. Straubmeier$^1$,
      L. Moser$^1$,
      N. Sabha$^1$,
      R. Buchholz$^1$, \\
      M. Zamaninasab$^{2}$, 
      K. Mu\v{z}i\'{c}$^{10}$, 
      J. Moultaka$^{11}$,
      J. A. Zensus$^{2}$ 
}

\address{
{1)}   I.Physikalisches Institut, Universit\"at zu K\"oln,
 Z\"ulpicher Str.77, 50937 K\"oln, Germany
\\
{2)} Max-Planck-Institut f\"ur Radioastronomie, 
             Auf dem H\"ugel 69, 53121 Bonn, Germany,
\\
{3)} Department of Astronomy, University of Maryland, College Park, MD 20742-2421, USA,
\\
{4)} University of California, Los Angeles, Department of Physics and Astronomy, USA,
\\
{5)} Massachusetts Institute of Technology, Kavli Institute for Astrophysics and Space Research, USA,
\\
{6)}	Department of Physics, University of Washington, Seattle, WA 98195-1560, USA ,
\\
{7)}Theoretical Astrophysics Center and Department of Astronomy,
University of California, Berkeley, CA 94720-3411, USA
\\
{8)} Instituto de Astrof\'isica de Andaluc\'ia (CSIC), Glorieta de la Astronom\'ia s/n, 18008 Granada, Spain,
\\
{9)} Astronomical Institute AV CR, Prague, Czech Republic,
\\
{10)} Department of Astronomy \& Astrophysics, 
              University of Toronto, 50 St. George Street, 
              Toronto, ON M5S 3H4, Canada 
\\
{11)} LATT, Universit\'e de Toulouse, CNRS, 14, Avenue Edouard Belin, 31400 Toulouse, France
}

\ead{eckart@ph1.uni-koeln.de}

\begin{abstract}
Based on Bremer et al. (2011) and Eckart et al. (2012) 
we report on simultaneous observations and modeling 
of the millimeter, near-infrared, and X-ray flare emission of the 
source Sagittarius A* (SgrA*) 
associated with the super-massive (4$\times$10$^6$\solm) 
black hole at the Galactic Center.  
We study physical processes giving rise to the variable 
emission of SgrA* from the radio to the X-ray domain.
To explain the statistics of the observed variability  
of the (sub-)mm spectrum of SgrA*, we use
a sample of simultaneous NIR/X-ray flare peaks
and model the flares using a synchrotron and SSC mechanism.
The observations reveal flaring activity in all wavelength bands 
that can be modeled as the signal from adiabatically expanding 
synchrotron self-Compton (SSC) components.
The model parameters suggest that either the 
adiabatically expanding source components
have a bulk motion larger than v$_{exp}$ or the expanding material 
contributes to a corona or disk, confined to the immediate 
surroundings of SgrA*.
For the bulk of the synchrotron and SSC models, we find 
synchrotron turnover frequencies in the range 300-400~GHz.
For the pure synchrotron models this results in densities of relativistic 
particles of the order of 
10$^{6.5}$cm$^{-3}$ and for the SSC models, the median densities are
about one order of magnitude higher.
However, to obtain a realistic description of the 
frequency-dependent variability amplitude of SgrA*,
models with higher turnover frequencies and 
even higher densities are required.
We discuss the results in the framework of possible deviations from 
equilibrium between particle and magnetic field energy. 
We also summarize alternative models to explain the broad-band
variability of SgrA*.
\end{abstract}

\section{Introduction}
\label{introduction}

Sagittarius~A* (SgrA*) is the closest super-massive black hole (SMBH)
and is the prime candidate to study the 
variability and spectral properties of accreting SMBHs.
Since it is at the center of the Milky Way at a distance of about 8~kpc, 
Doeleman et al. (2008, 2009) and  Fish et al. (2011) have
recently succeeded in obtaining structural information on 
event-horizon scales
through very long baseline interferometry (VLBI) 
at 1.3~mm wavelength.
SgrA* is strongly variable in the radio and millimeter wavelength regime
(Zhao et al. 2003; 
Mauerhan et al. 2005;
Eckart et al. 2008ac; Marrone et al. 2008; 
Li et al. 2009; Yusef-Zadeh et al. 2008, 2009).
It shows an 
inverted spectrum from the radio to the (sub-)millimeter 
domain (Falcke et al. 2000)
and displays order-of-magnitude flares in the infrared and 
X-ray domain (Baganoff et al. 2001, 2003; Genzel et al. 2003; 
Eckart et al. 2004, 2006a, 2010, Ghez et al. 2003, 2004). 
The sub-mm spectrum of SgrA* itself is rather unexplored 
owing to the difficulty of separating it from contributions
of the surrounding ''mini-spiral''  and the circum-nuclear disk (CND).

While the so-called sub-millimeter bump is attributed to
SgrA* and is thought to be due to relativistic, thermal electrons 
of a hot, thick, advection-dominated
accretion flow (Dexter et al. 2009, 2010; Narayan et al. 1995; Yuan et al. 2003), 
it is, however, currently unclear whether the radio spectrum is
due to contributions from  an accretion flow (Yuan et al. 2003), 
due to a jet (Falcke \& Markoff 2000), or a combination of both. 
The strong variability observed at radio to X-ray emission is most likely due to
synchrotron and/or synchrotron self-Compton (SSC) radiation.
Part of the radiation may be due to a single hot-spot or a multi-spot 
model in the mid-plane of the accretion flow
or an increased accretion rate
(see models by Broderick \& Loeb 2006 and Eckart et al. 2006b, 2011, Meyer et al. 2007, 
Yuan et al. 2008, see also Pech\'a\v{c}ek et al. 2008, 
Shcherbakov \& Baganoff 2010, Dexter et al. 2010, Zamaninasab et al. 2011).
Spiral arm models 
(e.g. Karas et al. 2007) or jet/jet-base models
(Falcke \& Markoff 2000, Markoff 2005, Markoff et al. 2007)
are also possible.
Another result from multi-wavelength observations of  SgrA* is that 
the process of adiabatic expansion of source components 
may be relevant (Eckart et al. 2006a, 2008abc, 2009, 2010, 2012,
Yusef-Zadeh et al. 2006a, 2008, 2009).
Here we assume that the sub-mm, NIR and X-ray  flux excursions
are physically related
(However, see a detailed discussion on this in section 5.3 in Eckart et al. 2012).
This expansion can explain the observed time lags between the
infrared/X-ray and millimeter emission peaks. 
The fact that one observes different time delays 
between NIR and sub-mm flares can be understood as a consequence of a 
spread in source sizes and synchrotron turnover frequencies.
The idea of adiabatic expansion is supported by the fact that there
is no conclusive observational evidence for sub-mm flares preceeding NIR events
and by the fact that it has been detected at radio cm-wavelengths
(Yusef-Zadeh et al. 2006ab).
Here we summarize results obtained for simultaneous flare emissions 
in the NIR and X-ray with implications to the observed radio variability.
In this summary we include the analysis of the NIR spectral index
under the assumption that especially for weak flares synchrotron losses are important.
A detailed description of the Synchrotron/SSC analysis is given in Eckart et al. (2012).
Throughout the article we use $S_{\nu} \propto \nu^{-\alpha}$.

\section{Observations}
\label{observations}
The analysis summarized in this paper is based on
Bremer et al. (2011) and Eckart et al. (2012) 
who have studied a sample of NIR H- and Ks-band flares 
and a sample of all X-ray flares (until 2010) that showed
within less than about 10 minutes a  near-infrared emission peak.
The detailed flare data are listed in Tab.3 by Eckart et al. (2012).
In this paper we use a consistent 
calibration based on the extinction value of $A_K$=2.46 (Sch\"odel et al. 2010).
For the L'-band flares (Dodds-Eden et al. 2009; Trap et al. 2011ab),
we used $A{_L'}$=1.23 (Sch\"odel et al. 2010) and extrapolated to the
Ks-band with a spectral index of +0.7$\pm$0.3 
(Hornstein et al. 2007, see also Bremer et al. 2011). 

\section{Interpretation}
\label{interpretation}

\subsection{The NIR spectral index}
\label{index}

Bremer et al. (2011) point out that the broad-band near-infrared 
spectral index is an essential quantity to investigate the 
physical mechanism that underlies the observed SgrA* flare emission.
The authors present a method to derive the NIR spectral index of SgrA*
between the H- and Ks-band from the statistics of the peak flux density 
of the SgrA* flares. 
They base their spectral index derivation on an unprecedentedly 
large time-base of about seven years of monitoring 
the infrared counterpart of SgrA*.
The underlying assumption is that for SgrA* in both NIR bands 
the same optically thin, dominant emission mechanism is at work 
and produces similar number distributions of flares.
The author could then cross-correlate the peak flare flux histograms 
and determine the expectation value of the H-Ks-band spectral index 
during the bright flare phases of SgrA*.
(For a detailed analysis of the possible contribution of stars within the
central arcseconds to the flux density at (of close to) the position of SgrA*
and the statistical analysis of K-band light curves see Sabha et al. 2012 and
Witzel et al. 2012).

Using this method, Bremer et al. (2011) can confirm that the 
spectral index for brighter flares is consistent with $\alpha$=+0.7 
$\left(S_{\nu} \propto \nu^{-\alpha} \right)$
which is indicative of optically thin synchrotron radiation. 
In addition the authors find a tendency for the weaker flares 
to exhibit a steeper near-infrared spectrum.
From this they can conclude that the distribution of spectral 
indices as a function of peak Ks-band flare flux density can 
successfully be represented by an exponential cutoff proportional to 
$exp[-(\nu/\nu_0)^{0.5}]$, were $\nu_0$ is a characteristic cutoff frequency.
Such a cutoff is expected if synchrotron losses are at work. 
If $\nu_0$ varies between the NIR and sub-mm wavelength domain,
and if the peak flare sub-mm flux density variation is about one Jansky,
then the model explains the observed spectral properties of SgrA* in the NIR.
For the model calculations we assume that the intrinsic optically thin 
spectral index of these synchrotron flares is $\alpha$=+0.7$\pm$0.3.
The spectral index cannot be substantially steeper in order not to violate
the MIR flux density limit (Sch\"odel et al. 2007, 2011).
For a peak flare flux density in the Ks-band of 7-6~mJy, the spectral index of
+0.7 results in a $\sim$1.5~mJy flux density difference with respect to the H-band.
This corresponds to a $\ge$3$\sigma$ quantity for the relative calibration 
uncertainty of about 10\% between the two NIR bands.
For the combined $\ge$30 flares per Ks- and H-band the flux density differences
due to the +0.7 spectral index is significant for the bulk of the detected flares.

\begin{figure}
\begin{center}
\includegraphics[width=15cm]{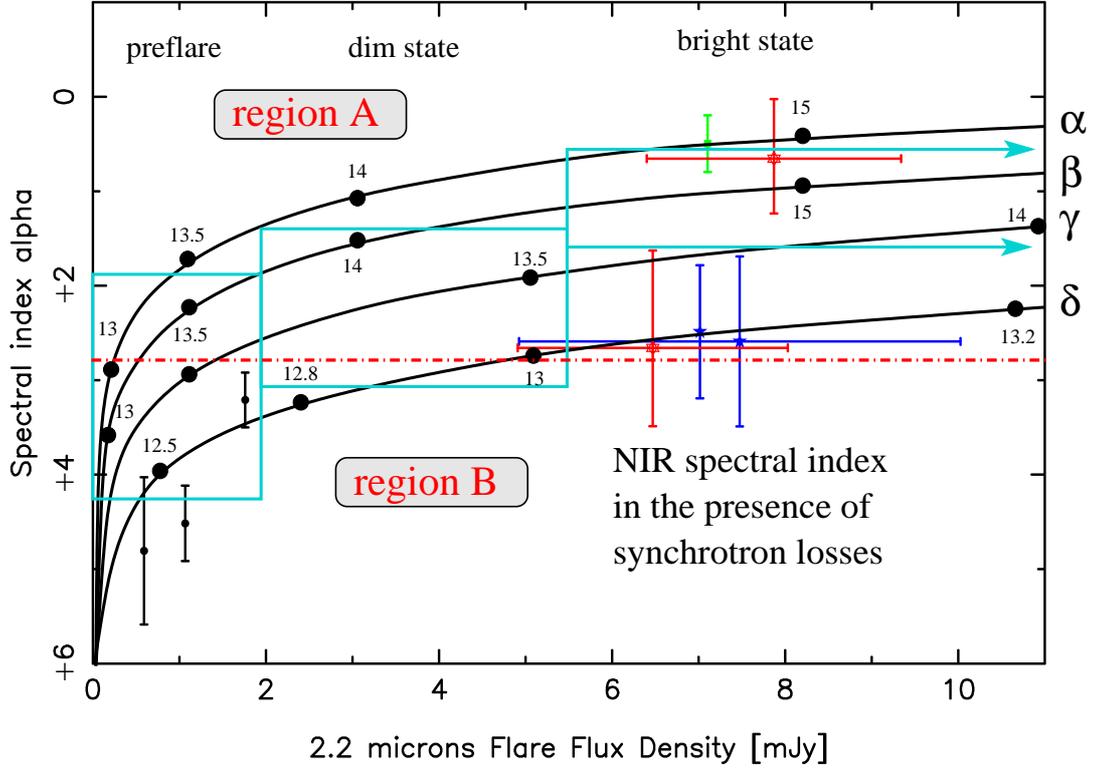}
\end{center}
\caption{
\small
The H-Ks-band spectral index versus the de-reddened Ks-band flux density
(Bremer et al. 2011).
The black dots are the logarithms of the cutoff frequencies and the 
black solid lines are theoretical curves $\alpha$ to $\delta$ explained in the text.
These curves are parameterized and labeled with the logarithm of the cutoff frequency.
The boxes and the regions A and B separated by a dashed-dotted line 
are explained in the text.
\label{fig:result2}
}
\end{figure}

Fig.~\ref{fig:result2} shows the relation between the H-Ks-band 
spectral index and the observed Ks-band peak flux density as obtained for 
a power law assuming an exponential cutoff due to synchrotron losses.
The thick solid black curves show model calculations 
that cover the expected spectral synchrotron indices  
and a 1~THz flare strength corresponding to the 
degree of variability observed at high frequencies.
Model $\alpha$ is based on a power law spectral index of 
$\alpha$=+0.4 and a flux of 0.071 Jy at 1~THz. 
As laid out in Eckart et al. (2012) 
we used for model $\beta$ a value of $\alpha$=+0.8 and 0.5~Jy,
for model $\gamma$ values of  $\alpha$=+0.6 and 0.76~Jy, and 
for model $\delta$ values of $\alpha$=+0.4 and 1.4~Jy.
Above a Ks-band flux density of 12~mJy, models 
$\alpha$, $\gamma$, and $\delta$
have flux densities above 8.5~mJy, and model $\beta$
has a 8.6~$\mu$m flux densities above 
the 22~mJy flux limit (Sch\"odel et al. 2007, Sch\"odel et al. 2011)
at which SgrA* has not yet been detected.

In Fig.~\ref{fig:result2} we also show
rectangular boxes that mark the flux densities and NIR 
spectral index ranges obtained from the broad-band NIR spectroscopy 
(Gillessen et al. 2006) from the H-/K'-imaging 
(Hornstein et al. 2007, see also Bremer et al. 2011).
The values that correspond to the blue boxes are given in Tab.4 of
Bremer et al. (2011).
In region A the sub-mm/NIR flares are brighter and have flatter 
synchrotron spectra. The synchrotron loss turnover frequencies 
are well above 10$^{13}$Hz.
In region B the sub-mm/NIR flares are weaker and steeper, with
synchrotron loss turnover frequencies 
are below 10$^{13}$Hz.

\subsection{NIR/X-ray flares}
\label{nirxrayflares}

In Eckart et al. (2012) we have described physical and statistical aspects 
of the relation between sub-mm and NIR/X-ray flare emission of SgrA*. 
For most synchrotron and SSC models, the authors find 
synchrotron turnover frequencies in the range of 300-400~GHz.
For the pure synchrotron models this gives densities of relativistic 
particles of the order of 10$^{6.5}$cm$^{-3}$.
For the pure SSC models, the median densities are
about one order of magnitude higher.
In order to obtain, however, a realistic description of the 
frequency-dependent variability amplitude of SgrA*,
we require models with substantially higher turnover frequencies and 
higher densities.

\subsubsection{Formalism \\}
\label{formalism}
\noindent
\vspace*{-4mm}
\\
Eckart et al. (2012) adopt the formulae previously presented 
by  Marscher (1983, 2009). For the case
of the Galactic Center, they find that the SSC X-ray flux density
$S_{X,SSC}$ (in $\mu$Jy),
magnetic field $B$ (in G), and column density of relativistic electrons 
$N_0$ (in cm$^{-3}$keV$^{-1}$) involved in the emission process 
can be described by

\begin{equation}
\label{eq1}
S_{X,SSC}=d(\alpha)\ln(\frac{\nu_2}{\nu_m})\theta^{-2(2\alpha+3)}\nu_m^{-(3\alpha+5)}S_m^{2(\alpha+2)}E_X^{-\alpha}\delta^{-2(\alpha+2)}~~,
\end{equation}

\begin{equation}
\label{eq2}
B=10^{-5}b(\alpha)\theta^4\nu_m^5S_m^{-2}\delta~~,
\end{equation}

\begin{equation}
\label{eq3}
N_0=n(\alpha)D_{Gpc}^{-1}\theta^{-(4\alpha+7)}\nu_m^{-(4\alpha+5)}S_m^{2\alpha+3}\delta^{-2(\alpha+2)}~~.
\end{equation}

\noindent
Here $d(\alpha)$, $b(\alpha$), and $n(\alpha)$ are dimensionless parameters (see Marscher 1983). 
For the case that they are all functions of $\alpha$, $D_{Gpc}$ is the luminosity distance in 
gigaparsecs and $E_X$ the X-ray photon energy in $keV$.
The underlying relativistic electron distribution is expressed as
$N=N_0~\exp(-p)$ with an electron power-law index $p$.

Eckart et al. (2012) then outline that the peak flux $S_m$, source size $\theta$, magnetic field $B$, and relativistic particle density $N_0$ can all be
expressed as power laws of the frequency $\nu$: 

\begin{equation}
S_m=\kappa_1 \nu_m^{-\alpha}~~,
\end{equation}
\begin{equation}
\theta = \kappa_2 \nu_m^{\zeta_1}~~,
\end{equation}
\begin{equation}
B=\hat{\rho} \nu_m^{\zeta_2}~~,
\end{equation}
and
\begin{equation}
N_0=\kappa_3 \nu_m^{\zeta_3}~~,
\end{equation}

with coefficients and exponents given in their paper.

\begin{figure}
\begin{center}
\includegraphics[width=15cm]{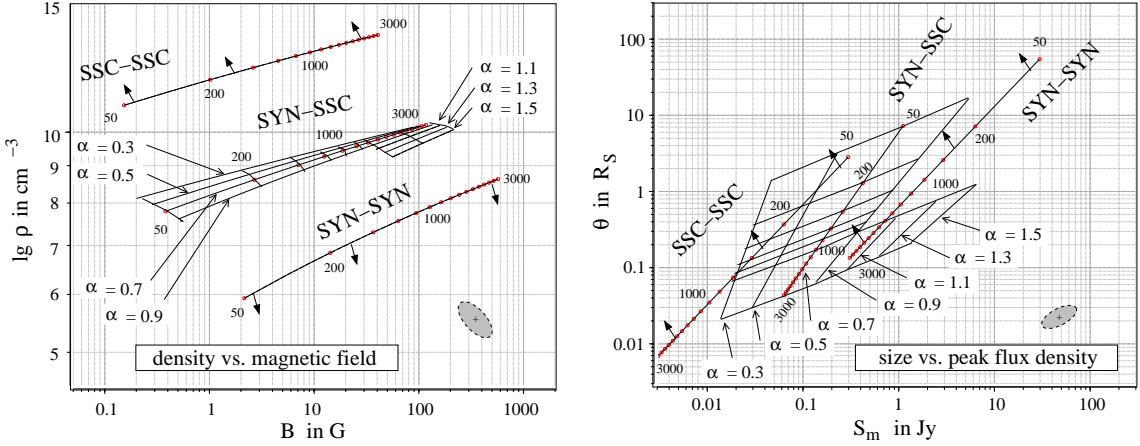}
\end{center}
\caption{
\small
Volume density of relativistic electrons versus magnetic field and
source size versus synchrotron peak flux density for the three
major source models SYN-SYN, SYN-SSC and SSC-SSC as described in the text and in Eckart et al. (2012).
The graphs describe a specific flare observed on 7 July 2004.
The ellipses represent uncertainties and
the thick arrows indicate the direction into which the
SSC-SSC and SYN-SYN model line will move if the synchrotron or self-Compton
limit is lowered.
The red circles are individual model data points connected with black solid lines.
Limits on the density, magnetic field, size and peak flux density $S_m$ result in a
spectral distribution of S$_m$ values that can be compared with the distribution of the observed data.
Details for the figure are given in the text and by Eckart et al. (2012). 
\label{result:monster}
}
\end{figure}

\subsubsection{Three cases of describing the spectra \\}
\label{threeclasses}
\noindent
\vspace*{-4mm}
\\
One can think of three different, basic cases in order to discuss 
the contributions of a single source emitting synchrotron (SYN) 
and synchrotron self-Compton (SSC) 
radiation in the radio, NIR and X-ray domains.
The three combinations are abbreviated as SSC-SSC, SYN-SYN and SYN-SSC.
The first label SSC-SSC indicates that both the
NIR and X-ray fluxes are due to SSC
scattered photons and less than 10\% are due to a pure SYN contribution.
The second label SYN-SYN indicates that a dominant contribution in these
two wavelength bands is due to a SYN contribution and again less than
10\% is due to a SSC contribution.
The third label SYN-SSC indicates that the NIR flux density is mainly
due to a SYN contribution and that the predominant amount of X-ray 
radiation is caused by a SSC contribution.
In the two cases of SYN-SYN and SSC-SSC, the spectral indices in the 
NIR and X-ray domains are the same (unless synchrotron losses are involved). 
The spectral index can be calculated from the flux densities in the
two bands. 
In Fig.~\ref{result:monster} we show the results of model calculations for one particular flare.
In the case of SYN-SSC, the graphs are labeled with the 
value of the optically thin SYN spectral index.
The model graphs are only shown as thin solid lines if the NIR and X-ray
flux densities are not contaminated by more than 10\% by the 
SSC or pure SYN contributions, respectively.
Mixed models in which both contaminating contributions are
higher than 10\% can be located 
between the SYN-SYN and SSC-SSC graphs.
The models also obey the MIR flux density limit 
(Sch\"odel et al. 2007, Sch\"odel et al. 2011; see Eckart et al. 2012 for details).

Models in Fig.~\ref{result:monster} were calculated for 
a cutoff frequency range between 50~GHz and 3~THz.
The figure shows that the typical ranges of 
relativistic electron densities for the SYN-SYN, the
SYN-SSC, and the SSC-SSC models are around 
$10^7$, $10^9$, and $10^{12}$ cm$^{-3}$, respectively.  
In general the relativistic electron density 
increases with increasing magnetic field
strength $B$, spectral index $\alpha$, and synchrotron cutoff frequency.
From SSC-SSC via SYN-SSC to SYN-SYN models 
the peak flux density $S_m$ and source size $\theta$ both increase.
This is also the case for decreasing 
cutoff frequency $\nu_m$ and spectral index $\alpha$.

For $\theta$$<$2~$R_S$, $S_m$$<2~Jy$, and $B<30~G$, a larger number of
NIR/X-ray flares can be described by SYN-SSC and SSC-SSC models, and cutoff 
frequencies of a few hundred GHz are preferred.
For the rejected SSC-SSC and the SYN-SYN models, the violation of the 
MIR flux density limit prohibits these models.
For the SYN-SSC, models can be rejected on the basis of 
excessive magnetic field strength $B$, source size $\theta$, and peak flux density $S_m$.

We consider the observed frequency-dependent variability amplitude ranges as
an important observable that needs to be explained by SgrA* source models.
Eckart et al. (2012) list in Tab.5  
the upper limits to $B$, $S_m$, and $\theta$, the lower limit to log($\rho$),
along with the best-fit $\chi^2$ value obtained using the given parameter limits.
For the SYN-SSC case, we used the full range of spectral indices  
to describe the sub-mm flux density variation of SgrA*.
With the restricted set of parameters in Tab.5 (Eckart et al. 2012) 
the models represent the 
observed frequency-dependent variability amplitudes quite well.
As explained in more detail in Eckart et al. (2012) 
for the SSC-SSC case, the degree of variability at high frequencies is too small
to reflect the measurements well.
For the SYN-SYN case, the degree of variability at low frequencies 
is much larger than observed. Also, most models are unable to provide an 
overall description of the SgrA* variability.
Under the assumption of a single source component we can summarize that
only the SYN-SSC model is capable in reproducing the
observed  SgrA* variability.

\begin{figure}
\begin{center}
\includegraphics[width=15cm]{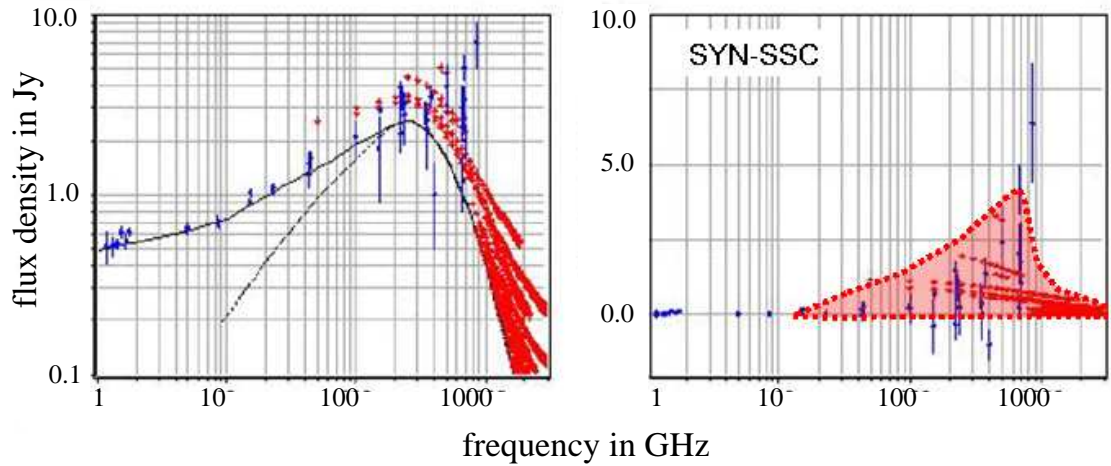}
\end{center}
\caption{
\small
Observed flux densities of Sagittarius~A* 
(Eckart et al. 2012) taken from the literature (blue) compared to
a combined model that consists of the fit given by Falcke et al. (2000),
Marrone et al. (2008) (black line),  and Dexter et al. (2010) 
(black dashed line). 
The spectrum of synchrotron self-absorption frequencies for the 
range of models discussed here is shown in red.
We show results for the SYN-SSC models corresponding to the parameters in 
Tab.5 by  Eckart et al. (2012).
The SYN-SSC is the preferred model and represents most closely 
the observed variability of SgrA*.
On the left, we show the complete model, on the right we show only the 
variable flux densities as observed (blue) and modeled (red).
\label{fig:result3}
}
\end{figure}

\subsubsection{Equipartition \\}
\label{equipartition}
\noindent
\vspace*{-4mm}
\\
The analysis by Eckart et al. (2012) has shown that 
for a realistic description of the observed frequency-dependent 
variability amplitudes of SgrA* the single source component descriptions 
require 
high turnover frequencies (rather a few 100~GHz than a few 10~GHz) and at least 
densities that are 10 to 100 times
the mean density of the larger scale accretion flow.
The exact properties of matter close 
to the SgrA* SMBH are not constrained very well by existing observations.
However, the accretion rate and hence the central density may  
be much higher than the limits derived from Faraday rotation.
Also the magnetic field equipartition fraction and the
bias field strength in the case of magnetic field reversals 
are not well constrained 
(see Marrone et al. 2007 and Igumenshchev et al. 2003 for details).
While equipartition in the larger scale SgrA* accretion flow 
appears to be justified, deviations from equipartition may 
occur close to the midplane or temporary accretion disk of SgrA*.
Such deviations are likely to occur as a consequence of plasma 
instabilities in the immediate vicinity of SgrA* as two-stream instabilities of 
collisionless shocks.

The two-stream instability is known as a common instability in plasma physics. 
One of the ways to induce it involves a high energetic particle stream injected 
into a plasma. Another way is by creating conditions in which 
different plasma species (ions and electrons) have different drift velocities. 
The energy contained and released from the particles can then lead to 
plasma wave excitation.
The importance of the two-stream instability e.g. in Gamma-Ray Burst Sources (GBR)
is highlighted by Medvedev \& Loeb (1999). 
Through the generation of magnetic fields in 
in a relativistic shock, in this case one expects deviations from
equipartition between magnetic field and particle energy by a factor of up to $10^5$.
The effect of beam-plasma instabilities on accretion disk and the associated
flares are discussed in
Krishan, Wiita \& Ramadurai (2000) and
Keppens, Casse \& Goedbloed (2002).
The importance of acceleration of particles by 1st and 2nd order Fermi processes 
in the framework of relativistic jets and accretion discs around SMBH
is highlighted by e.g. Kowal et al. (2011) and Nishikawa et al (2011).

Particles may interact with each other not through Coulomb collisions but through 
collective behavior within plasma waves, i.e. collisionless shocks.
In these shocks, the transition from pre-shock to post-shock states occurs on a 
length scale that is much smaller than the mean free path for particle collisions.
The Weibel instability (Weibel 1959) represents such a collective response. 
It can occur both in magnetized and initially non-magnetized 
low-temperature relativistic flows.
In the latter they work apparently best (see details in the review by Bykov \& Treumann 2011). 
Another way to induce them is by having flows parallel to an ambient magnetic field - as 
e.g. in case of a toroidal magnetic field structure within the mid-plane of an accretion flow
or an accretion disk.
For initially non-magnetic and for magnetic shocks the average
magnetic energy remains at least about 15-20\% below the equipartition at the shock.
Magnetic fields induced by Weibel instabilities may form vortices, and 
3D PIC (particle in cell) simulations indicate that they may result in 
an extended downstream region with small-scale magnetic fields.
In the framework of accretion disks they might then be regarded as single or multiple hot spots.
\\
\\
{\it The case of SgrA*:~~}
Following Homan et al. (2006) and Readhead (1994) the intrinsic brightness 
temperature, $T_{int}$, is related to the equipartition brightness temperature $T_{eq}$ at 
energy balance by the expression
$T_{int}=\eta^{1/8.5}T_{eq}$, where $\eta=u_p/u_B$ is the ratio of the energy
densities of the radiating particles, $u_p$, to the energy density of the 
magnetic field, $u_B$. 
The exponent of $\eta$ assumes
that the radiating particles follow a power-law Lorentz distribution
with index $p=2.5$ for $N_{\gamma}d\gamma = K \gamma^{p} $ and $p=2\alpha+1$.
This value of $p$ corresponds to a spectral index of $\alpha=0.75$ 
($S\propto \nu^{-\alpha}$) in the optically thin parts of a jet or outflow.

Recent and ongoing VLBI observations at 1.3~mm (Doeleman et al. 2008, 2009;
Fish et al. 2011) determined a lower limit to the brightness temperature 
of the VLBI source SgrA* at 230~GHz of $T_b = 2 \times 10^{10}$~K.
This is close to the equipartition brightness temperature
of $T_{eq}=5 \times 10^{10}$~K that has been found by 
Readhead (1994) for a sample of accreting super-massive black holes.
So we may assume that the mm-radiation source SgrA* 
is close to equipartition in its more quiescent phases.
The intrinsic brightness temperature of SgrA* is then
limited either by the inverse Compton limit ($T_{int} \le 10^{12}$ K; 
Kellermann \& Pauliny-Toth 1969, Pauliny-Toth \& Kellermann 1966)
or by the possibility
that the emitting source is near equipartition to begin with so that 
$T_{int}= T_{eq}$ (Singal \& Gopal-Krishna 1985; Singal 1986; Readhead 1994).
We measure $T_{obs} = \delta T_{int}$, $\delta$ being the relativistic 
boosting factor and for SgrA* we can assume $\delta$$\sim$1 
(details in Eckart et al. 2012).
In the sub-mm regime SgrA* can easily show flares that are 30\% to 100\% 
in excess over the low state (which may be subject to longer variability as 
well; see  Dexter et al. 2010).
This already implies significant deviations from equipartition 
with values for $\eta$ in the range of 10 to 350.
As outlined above, such an imbalance may be the result of an injection or acceleration
of particles generated through plasma instabilities close to the SMBH and possibly
at the footpoint of a jet or outflow.

\subsection{Polarization}
\label{polarization}

The radio/NIR/X-ray flares (e.g. Fig.3) of increased flux last for about 100 min and are usually
accompanied by variations in the polarized emission
(e.g. Eckart \& Genzel 1996, Ghez et al. 1998,2004 , Sch\"odel et al. 2002,
Baganoff et al. 2001, 2003, Eckart et al. 2004, 2006b, 2008ab, Nishiyama et al. 2009,
Zamaninasab et al. 2010, Kunneriath et al. 2010).
This time scale fits well with the mm-wave $\sim$hourly timescales and 
structure function knees at 30-100 min found on individual occasions 
by Mauerhan et al. (2005) and Yusef-Zadeh et al. (2011).
The mm/sub-mm wavelength linear polarization of SgrA* is just below 10\%
and variable in magnitude and position angle (PA) on timescales down to 
a few hours
(Bower et al. 1999a).
At 227 and 343 GHz Marrone et al. (2007) have determined an almost constant rotation measure
of about -5.6$\times$10$^{-5}$ rad m$^{-1}$. The mean intrinsic PA of 167$^o$$\pm$30$^o$ 
limits with its uncertainties, the accretion rate fluctuations to 25\%. The accretion
rate is limited to $<$2$\times$10$^{-7}$ M$_{solar}$ yr$^{-1}$ if the magnetic field
is near equipartition, ordered, and
largely radial, or has values of $<$2$\times$10$^{-9}$ M$_{solar}$ yr$^{-1}$
if there is a sub-equipartition, disordered, or toroidal field.  
Marrone et al. (2007) also find that 
the rather constant mean intrinsic PA is probably originating
in the sub-millimeter photosphere of SgrA*, rather than arising from rotation 
measure changes.  
Circular polarization has also been detected for SgrA* (~0.5\% at cm-wavelengths), 
with a rising polarization fraction from 1.4 to 15~GHz 
(Bower et al. 2002, Bower et al. 1999abc; Sault \& Macquart 1999) and further 
to 345 GHz (~1.2\% at sub-mm wavelengths; Munoz et al. 2012). 
This expected circular polarization at mm/sub-mm wavelengths
(Beckert \& Falcke 2002; Beckert 2003, Tsuboi et al. 2003; Bower et al. 2003ab, 2005)
is most likely due to Faraday conversion of linear polarization which is also 
measured with increasing intensity towards high frequencies (Munoz et al. 2012).
Hourly variable sub-mm emission (Marrone et al. 2006, Mauerhan et al. 2005, 
Miyazaki et al. 2004) and a very
compact sub-mm VLBI structure (Doeleman et al. 2008/9) possibly point towards plasma blobs on
relativistic orbits close to the event horizon. Hence, simultaneous observations  
of the variability at mm/sub-mm wavelengths are fundamental for constraining the radiative
processes. Contributions to the flare emission from red noise processes are likely to be important
(see Do et al. 2009, Eckart et al. 2008a). 
However, Zamaninasab et al. (2010, 2011) have demonstrated that highly polarized
NIR sub-flares (on time scales of $\sim$20 min) are statistically significant compared to the 
randomly polarized red-noise.
In addition to intensity variations due to relativistic boosting, the PA is rotated
in the strong gravitational field of the SMBH. In these cases we can therefore estimate spin
and spin-vector orientation of the SMBH (i.e. the accretion disk) independently from both simultaneous
NIR and mm-polarization data in comparison to disk and jet models that we
generate using a relativistic 2D- and 3D-disk codes 
(Dovciak, Karas, \& Yaqoob 2004 and Bursa et al. 2012, this volume; 
running in Prague and Cologne). 
This code is capable of handling polarized emission and relativistic radiative transfer. It allows us to
describe an orbiting and gradually evolving spot on different trajectories (in a disk or jet) near a
rotating black hole including optical depth effects.

\subsection{Alternative models}
\label{alternatives}

In the SSC-SSC, SYN-SSC, and SYN-SYN models discussed above (see section~\ref{threeclasses})
a single population of relativistic electrons is responsible
for the emission at different wavelengths. One can think of more complex 
models in which different populations of thermal or relativistic electrons
are responsible for the emission.
Here we briefly discuss two of these models.
In both cases the SYN and SSC contributions have to be carefully tuned such that
they contribute predominantly only in one frequency domain and cannot be
attributed to the three cases mentioned above.

\subsubsection{Double-SYN \\}
\label{doubleSYN}
\noindent
\vspace*{-4mm}
\\
If we assume that SgrA* is surrounded by a magnetized corona
and - at least temporarily - by an accretion disk 
then the accreting medium is not uniform, and it may show 
inhomogeneities due to plasma instabilities operating in the weakly 
magnetized accretion flow. 
Such a dilute magnetosphere may contain individual 
magnetized source components which may lead to 
flares as they approach the black hole horizon 
(Galeev, Rosner \&  Vaiana, 1979, 
and subsequent works on patchy coronae).
These two components may both be the source of
synchrotron emission\footnote{We do not see the necessity to propose 
a double-SSC model, given that substantial synchrotron emission as well as
a high density of relativistic electron is required for both components.}. 
They may have different sizes, magnetic fields
and particle densities and - more importantly - they may
exist close to each other and may even be connected causally,
e.g. outflows and the possibly back flowing envelopes or
cocoons, or they may represent a jet or outflow and its base.
In this double-SYN case the more compact or denser component 
could be responsible for the synchrotron X-ray contribution and the 
large and more tenuous component could be responsible for the 
mm/NIR synchrotron contribution.

\subsubsection{External Compton scattering \\}
\label{externalCompton}
\noindent
\vspace*{-4mm}
\\
The region responsible for the infrared emission may be different from that of the 
mm-emission and the mm-emission region may be optically thin for near-IR photons. 
Then the relativistic electrons in this source component are available to 
up-scatter near-IR seed photons to X-ray energies (Yusef-Zadeh et al. 2009). 
Since the scattering source is not identical in this case with the seed photon
generating source, one may attribute this mechanism to the external inverse Compton 
scattering (ICS) mechanism in contrast to the synchrotron self Compton (SSC) excess.
In order to allow for a high efficiency for inverse Compton scattering, the density of the
scattering electrons and the scattered seed photons must be high. (For non-thermal
sources this is therefore usually best done by the source also responsible for the
seed photons i.e. SSC.) 
Again the individual synchrotron and synchrotron self Compton contributions have
to be well tuned such that the pure inverse Compton contribution remains dominant in
the X-ray domain.
This whole scenario may be complicated even further if one considers that the mm-wavelength
emitting cloud may be optically thick and occults the quiescent or flaring emission 
from SgrA*  (Yusef-Zadeh et al. 2010).

However, if one accepts the statement that the mm-wavelength flares are physically associated 
with the NIR/X-ray flares, then the observed adiabatic expansion velocity of the
radio components (typically 0.01~c) and the observed radio/NIR/X-ray flare length of
20 minutes to one hour give a minimum size of the order of 0.3 to 1 $R_s$ (Schwarzschild radii).
However this is a typical size of an emitting source component close to the SMBH, 
such that sources - emitting at different wavelengths at or below this size - 
have to be treated as a single component (implying SSC rather than ICS) 
than two separate sources. 
Alternatively, one is forced to treat the emitting regions as independent regions
and treat them in a fully stochastical description.
These senarii become rapidly unphysical, since it is difficult to have a source 
only emitting in a single spectral domain without a significant contribution in
neighboring wavelength domains.

\subsubsection{Fermion and Boson balls \\}
\label{fermionboson}
\noindent
\vspace*{-4mm}
\\
There is the attempt to explain very massive objects at the centers of galaxies 
by the concept of Fermion or Boson balls.
The mere fact that we observe strong variability from SgrA* may already
speak for a rather stabile than a fragile mass concentration at the center.
A universal Fermion ball solution for compact galactic nuclei
can be excluded by the results from stellar orbits near SgrA* 
(see references in Eckart et al. 2008abc, 2010).
An alternative explanation is that of a massive Boson star
(Torres, Capozziello \& Lambiase 2000, Lu \& Torres 2003 and references 
therein). This explanation is severely challenged by the general agreement between 
the measured polarized flare structure and the theoretical predictions 
(e.g. Eckart et al. 2006b, Zamaninasab et al. 2011).
If the indicated signatures of orbiting matter (very close to the event horizon) 
are considered, then also a stationary Boson star may be excluded
as an alternative solution for SgrA*
(Eckart et al. 2006b and 
see also result for the nucleus of MCG-6-30-15 by Lu \& Torres 2003).
For these objects one expects the orbital periods to be larger than those 
inferred from modeling the NIR polarized flares.
Especially ongoing accretion and star formation
in the central cluster results in requirements
to keep Fermion or Boson balls stable, which are most likely not met.
A similar result is reached by Miller (2006) 
and Munyaneza \& Biermann (2005)
who discuss
constraints on alternatives to super-massive black holes
and their growth of super-massive black holes in galaxies.

\subsubsection{Stars, Planets and Asteroids \\}
\label{StarsPlanetsAsteroids}
\noindent
\vspace*{-4mm}
\\
The models for SgrA* presented above are based on descriptions derived 
for non-thermal sources associated with both stellar black holes and 
low- and high-luminosity accreting super-massive black holes in nuclei of external host galaxies.
Since the luminosity of SgrA* is rather low compared to other SMBH sources and 
since it is the closest of such sources that we can study, there are attempts to explain 
the flare emission of SgrA* via different mechanisms.
There are two major recent models: The first one involves a population of asteroids and planets
that are disrupted by tidal forces exerted on them by SgrA*
(Zubovas, Nayakshin \& Markoff 2011).
The second uses the coronal radiation of a cusp of spun-up stars to explain 
the X-ray luminosity of SgrA* (Sazonov, Sunyaev \& Revnivtsev 2012).
It needs to be investigated if both of these models can explain the full body of
observations of SgrA* and if they can be tested by further observations.

\section{Summary and Conclusion}
\label{summary}
Eckart et al. (2012) have shown that the bulk of synchrotron and 
SSC models applicable to SgrA* have preferred synchrotron turnover 
frequencies in the range of a few hundred GHz.
For the pure synchrotron models, the densities of relativistic 
particles are of the order of 10$^{6.5}$cm$^{-3}$.
For models involving SSC contributions
the median densities are an order of magnitude higher.
These values are quite comparable to those
quoted for the accretion flow toward SgrA*
(e.g.  Yuan et al. 2003, 2004, Yusef-Zadeh et al. 2006ab).

For a realistic description of the observed frequency-dependent 
variability amplitudes of SgrA* higher densities and turnover 
frequencies are also required (details in Eckart et al. 2012).
Higher densities are neither unreasonable nor out of reach.
The accretion rate and hence the central density may  
be much higher than the limits derived from Faraday rotation
(Bower et al. 2003, Bower 2003, Marrone 2006a, Marrone et al. 2006abc, 
and discussion by Marrone et al. 2007).
In particular, the magnetic field equipartition fraction as well as the
bias field strength in the case of magnetic field reversals - that 
can be expected for a turbulent flow - are unknown.  
Variations in the magnetic field structure as well as the field strength
with respect to equipartition (Marrone et al. 2007, Igumenshchev et al. 2003)
are likely to result in higher densities in the immediate vicinity of 
the central SMBH.
While SgrA* may be close to equipartition, it is quite likely that 
at least during flux density events associated with very strong X-ray flares
and the related NIR and (sub)mm-flux excursions,
SgrA* may go through phases in which it is off equipartition.

Currently, it is difficult to decide which of all the models mentioned 
here describe the dominant contributors to the flare activity of SgrA*,
or which combinations of effects are most likely or if one or some of the 
mechanisms can be excluded entirely.
A key appears to be the restrictions imposed onto the models by the frequency
dependant flux density variability or polarization properties. 
Therefore, the simultaneous flux density and
 polarization monitoring observing campaigns are essential.  
Any model (or combinations of those) must be able to explain these 
observations.
In addition, it is essential, that high angular resolution observations in the 
NIR (interferometry with long baselines) and radio domain (mm- and sub-mm-VLBI)
are being performed. These observations will help to identify the orbital motion 
around the SMBH or motion along a jet/outflow.
\\
\\
{\bf Acknowledgements: }
N. Sabha is member of the Bonn Cologne Graduate School (BCGS) for
Physics and Astronomy supported by the Deutsche Forschungsgemeinschaft.
M. Valencia-S. is member of the International Max-Planck Research School
(IMPRS) for Astronomy and Astrophysics at the Universities of Bonn and Cologne
supported by the Max Planck Society.
F. Baganoff was supported by NASA through Chandra Award 
Number GO9-0101X and SAO Award Number 2834-MIT-SAO-4018.
Part of this work was supported by the COST Action MP0905: 
Black Holes in a violent Universe and PECS project No. 98040.
We are grateful to all members
of the NAOS/CONICA, ESO PARANAL, and APEX team.  
M. Garc\'{\i}a-Mar\'{\i}n is supported by the German 
federal department for education and research (BMBF) under 
the project number 50OS1101.
R. Sch\"odel acknowledges support by the Ram\'on y Cajal program, 
by grants AYA2010-17631 and AYA2009-13036 of the Spanish Ministry of
Science and Innovation, and by grant P08-TIC-4075 of the Junta de
Andaluc\'ia.
Ongoing CARMA development and operations are supported by the 
National Science Foundation under a cooperative agreement,
and by the CARMA partner universities.

\end{document}